\documentclass[11pt]{article}
\usepackage[]{fontenc}
\usepackage[latin1]{inputenc}
\usepackage{graphics}

\makeatletter

\providecommand{\LyX}{L\kern-.1667em\lower.25em\hbox{Y}\kern-.125emX\@}

\usepackage[]{fontenc}
\usepackage[latin1]{inputenc}
\usepackage{graphics}

\makeatletter

\usepackage[]{fontenc}
\usepackage[latin1]{inputenc}

\makeatletter

\usepackage{amssymb}

\usepackage[]{fontenc}
\usepackage[latin1]{inputenc}
\usepackage{geometry}
\usepackage{setspace}
\usepackage[]{fontenc}
\usepackage[latin1]{inputenc}
\usepackage{graphics}
\usepackage{graphicx}

\makeatletter
\providecommand{\LyX}{L\kern-.1667em\lower.25em\hbox{Y}\kern-.125emX\@}
\makeatother

\makeatother

\makeatother

\makeatother

\begin{document}

{\par\centering {\huge Multi-channel pulse dynamics in a stabilized Ginzburg-Landau
system} \par}

\vspace{12mm}

{\par\centering {\large H.E. Nistazakis\( ^{1} \), D.J. Frantzeskakis\( ^{1} \),
J. Atai\( ^{2} \), B.A. Malomed\( ^{3} \), N. Efremidis\( ^{4} \), and K.
Hizanidis\( ^{4} \)} \vspace{3mm} \par}

{\par\centering \( ^{(1)} \)\textit{Department of Physics, University of Athens,
Panepistimiopolis, 15784 Athens, Greece}\\
\( ^{(2)} \)\textit{School of Electrical and Information Engineering, The University
of Sydney, Sydney, NSW 2006, Australia}\\
\vspace{1.5mm} \par}

{\par\centering \( ^{(3)} \)\textit{Department of Interdisciplinary Studies,
Faculty of Engineering, Tel Aviv University, Tel Aviv 69978, Israel}\\
\( ^{(4)} \)\textit{Department of Electrical and Computer Engineering, National
Technical University of Athens, 15773 Athens, Greece} \vspace{15mm} \par}

{\par\centering \textbf{\large Abstract} \par}

{\par\centering \vspace{5mm}\par}

We study the stability and interactions of chirped solitary pulses in a system
of nonlinearly coupled cubic Ginzburg-Landau (CGL) equations with a group-velocity
mismatch between them, where each CGL equation is stabilized by linearly coupling
it to an additional linear dissipative equation. In the context of nonlinear
fiber optics, the model describes transmission and collisions of pulses at different
wavelengths in a dual-core fiber, in which the active core is furnished with
bandwidth-limited gain, while the other, passive (lossy) one is necessary for
stabilization of the solitary pulses. Complete and incomplete collisions of
pulses in two channels in the cases of anomalous and normal dispersion in the
active core are analyzed by means of perturbation theory and direct numerical
simulations. It is demonstrated that the model may readily support fully stable
pulses whose collisions are quasi-elastic, provided that the group-velocity
difference between the two channels exceeds a critical value. In the case of
quasi-elastic collisions, the temporal shift of pulses, predicted by the analytical
approach, is in semi-quantitative agrement with direct numerical results in
the case of anomalous dispersion (in the opposite case, the perturbation theory
does not apply). We also consider a simultaneous collision between pulses in
\emph{three} channels, concluding that this collision remains quasi-elastic,
and the pulses remain completely stable. Thus, the model may be a starting point
for the design of a stabilized wavelength-division-multiplexed (WDM) transmission
system.\\
PACS: 42.81.Dp, 42.65.Tg, 42.81.Qb

\newpage

\section{Introduction}

It is commonly known that complex cubic Ginzburg-Landau (CGL) equations constitute
a class of universal models for the description of pattern formation in various
nonlinear dissipative media \cite{ref1}. Equations of the CGL type are also
frequently used in nonlinear fiber optics, to describe the formation, stability,
and interactions of solitary pulses (SPs). CGL equations with constant coefficients
apply to a long nonlinear optic-fiber link if the pulses in it are broad enough,
so that the corresponding dispersion length is essentially larger than the amplification
spacing. In this case, the periodically placed amplifiers and filters (which
limit the gain to a relatively narrow spectral band) may be considered in the
uniformly-distributed approximation, neglecting their discreteness \cite{ref2}.

The single-component CGL equation with the cubic nonlinearity possesses a well-known
exact solitary-pulse solution \cite{ref3}, which includes an internal chirp
(phase curvature across the pulse). However, a fundamental drawback of this
solution is that it is \emph{unstable}, as the zero solution to the cubic CGL
equation, i.e., a background on top of which the pulse is built, is unstable
due to the presence of linear gain in the equation. Development of physically
realistic models in which solitary pulses are fully stable is a problem of an
obvious interest in its own right, and it also has profound importance for fiber-optic
communications (see the book \cite{ref2} and, for instance, a recent work \cite{ref4}),
as well as for the design of soliton-generating fiber-loop lasers \cite{ref5}.
In the context of optical telecommunications, an issue of fundamental significance
is the development of adequate models for multi-component systems, corresponding
to a wavelength-division-multiplexed (WDM) multi-channel scheme implemented
in the optical fiber. An objective is then to design a system supporting pulses
in all the channels, which must be stable against small perturbations and mutual
collisions (see, e.g., books \cite{ref6}, \cite{ref7} and a recent paper \cite{ref8}).

A single-channel system, which may suppress the instability of the zero solution,
simultaneously allowing for the existence of stationary pulses and thus opening
way for them to be stable, was proposed and studied by means of analytical perturbative
methods in Ref. \cite{ref9}, and then tested by direct simulations in Ref.
\cite{ref10}. In this system, the CGL equation is linearly coupled to an additional
dissipative equation, which is a linear one in the most fundamental and physically
relevant version of the model \cite{ref10,pla1}. In the context of optical
fibers, the system may be realized as a \emph{dual-core} fiber, in which an
active core carries the linear gain, filtering, temporal dispersion, and Kerr
nonlinearity, while the parallel-coupled core is lossy, its intrinsic nonlinearity,
dispersion, and filtering being negligible. It has recently been shown \cite{ref11}
that this model may describe transmission of fully stable optical solitary pulses
with an internal \emph{chirp} (intrinsic phase structure) in an indefinitely
long fiber-optic communication link.

In this paper, we study the stability and collisions of chirped solitary pulses
in a system of nonlinearly coupled CGL equations stabilized by means of the
aforementioned scheme, i.e., by linearly coupling each CGL equation to its own
linear dissipative counterpart. In its simplest versions that are considered
in this work, the model includes two or three nonlinearly coupled stabilized
subsystems, which is a prototype of a \emph{stabilized WDM system}. The most
interesting issues, on which we focus in this work, are the stability of SPs
in this system and collisions between them (including a simultaneous collision
between three pulses belonging to three channels). Besides the obvious relevance
to optical telecommunications, the obtained results are of interest in their
own right, demonstrating a new type of stable traveling pulses and collisions
between them in a generalized (multicomponent) CGL system.

The paper is structured as follows. In section 2, we give a detailed formulation
of the model with two channels, and exact solutions for SPs in each channel.
In section 3, we develop an analytical approach to the collision problem, based
on the perturbation theory. In particular, a prediction for position shifts
of the pulses in the case of a quasi-elastic collision is obtained in a fully
analytical form. In section 4, results of systematic direct simulations of the
collisions are displayed for both inelastic and quasi-elastic cases; in the
latter case, the analytical predictions are found to be in good agreement with
the numerical results (in a parametric region where the perturbation theory
applies). In the same section, a region in the model's parametric space is identified,
in which the pulses are \emph{fully} stable, i.e., against both small perturbations
and mutual collisions. A generalization for a three-channel model is briefly
considered in section 5, with a conclusion that the pulses are also stable against
simultaneous collision of three of them. The paper is concluded by section 6.

\section{The two-channel model and exact solutions for the pulses}

\subsection{The model}

The simplest version of the model describes the propagation of two waves, \( u \)
and \( v \), carried by two different wavelengths in the active core of a dual-core
optical fiber. The waves interact through the cross-phase modulation induced
by the Kerr effect in the active core. The fields \( u \) and \( v \) are
assumed to be linearly coupled to two other fields, \( \phi  \) and \( \psi  \)
respectively, which propagate in the passive core placed parallel to the active
one. In fact, it is not necessary to assume that all the long fiber-optic link
has a dual-core structure; instead, it is sufficient to have short segments
of the dual-core fiber periodically installed into the link. Then, in the same
uniformly-distributed approximation which was mentioned above in relation to
amplification and filtering, we may consider an \emph{effectively homogeneous}
dual-core fiber link. In fact, this approximation was already applied to the
single-channel dual-core model in Ref. \cite{ref12}.

Thus, the model is based on the following system of normalized equations governing
the propagation of the four above-mentioned electromagnetic waves in the two
linearly coupled fiber cores: \begin{eqnarray}
i\left( u_{z}+cu_{t}\right) +\left( \frac{1}{2}D-i\right) u_{tt}-iu+\left( |u|^{2}+\sigma |v|^{2}\right) u & = & K\phi ,\label{eq1} \\
i\left( v_{z}-cv_{t}\right) +\left( \frac{1}{2}D-i\right) v_{tt}-iv+\left( |v|^{2}+\sigma |u|^{2}\right) v & = & K\psi ,\label{eq2} \\
i\left( \phi _{z}+\delta \cdot \phi _{t}\right) +i\Gamma \phi  & = & Ku,\label{eq3} \\
i\left( \psi _{z}-\delta \cdot \psi _{t}\right) +i\Gamma \psi  & = & Kv,\label{eq4} 
\end{eqnarray}
 where the complex electric field envelopes \( u \), \( v \), \( \phi  \),
and \( \psi  \) are functions of the propagation distance \( z \) and retarded
time \( t \) which are defined in the usual way \cite{ref2}. In the active
core, which is equipped with the amplification and filtering, the fields \( u \)
and \( v \) obey Eqs. (\ref{eq1}) and (\ref{eq2}), that include the nonlinear
self-phase modulation, linear gain, and effective filtering (the latter term
is formally tantamount to diffusion in the \( t \)-space). Coefficients in
front of the terms in Eqs. (\ref{eq1}) and (\ref{eq2}) accounting for these
three basic effects are all normalized to be equal to 1. On the other hand,
the dispersion coefficient \( D \) is explicitly present in the equations,
\( D>0 \) and \( D<0 \) corresponding, respectively, to anomalous and normal
dispersion. The field envelopes \( u \) and \( v \) in the active core are
nonlinearly coupled to each other, interacting through the cross-phase modulation
(XPM) induced by the Kerr effect, which gives rise to the XPM coupling coefficient
\( \sigma =2 \) in Eqs. (\ref{eq1}) and (\ref{eq2}). The fields \( u \)
and \( v \) are linearly coupled, through the corresponding coefficient \( K \),
to their counterparts \( \phi  \) and \( \psi  \) in the linear dissipative
core, which is characterized by a loss coefficient \( \Gamma  \). Finally,
the parameters \( c \) and \( \delta  \) in Eqs. (\ref{eq1})-(\ref{eq4})
stand for the inverse-group-velocity differences between the co-propagating
waves in the active and passive (lossy) cores.

A numerical value of \( c \) (assuming that \( \delta =c \)) will play an
important role below. To estimate the value that is of practical interest, we
note that the difference in the inverse group velocity is simply related to
the frequency separation \( \Delta \omega  \) between the channels, \( c=\beta \Delta \omega  \),
where \( \beta =k^{\prime \prime } \) is the standard physical dispersion coefficient
\cite{ref2,ref6}. The frequency difference can be expressed in terms of the
wavelength separation \( \Delta \lambda  \), \( \Delta \omega =-\left( 2\pi nc_{0}/\lambda ^{2}\right) \Delta \lambda  \),
where \( \lambda  \) is the wavelength proper, \( c_{0} \) is the light velocity
in vacuum, and \( n \) is the refractive index. For applications, the case
of interest is the one with the pulse's temporal width \( \sim 10 \) ps \cite{ref2,ref6,ref7}
and the wavelength separation between \( 1 \) nm and \( 0.1 \) nm \cite{ref7}.
Using these values, and typical values of the dimensional parameters for which
the present model provides for the best stability of the pulses (see Eqs. (\ref{eq13})
and (\ref{eq14}) below), an estimate similar to that presented, e.g., in Ref.
\cite{ref13} shows that relevant dimensionless values of \( c \) belong to
an interval \begin{equation}
\label{20}
c\sim 20-200\, .
\end{equation}
 Numerical results will be presented for this region of the values of \( c \)
(see Figs. 6 and 7 below).

\subsection{The linear spectrum}

Before proceeding to the analysis of the full system, it is relevant to consider
its linear spectrum in the dissipationless limit, i.e., in the case when the
gain, filtering, and loss terms are dropped in Eqs. (\ref{eq1})-(\ref{eq4}).
In the linear limit, the two subsystems (\( u,\phi  \)) and (\( v,\psi  \))
are decoupled, and, looking for the solution to the linearized equations in
the ordinary form \( \sim \exp \left( ikz-i\omega t\right)  \), one arrives
at the following dispersion relations between the propagation distance \( k \)
and frequency \( \omega  \): \begin{equation}
\label{q}
q=\frac{1}{4D}\left\{ \left( c-\delta \right) ^{2}-D^{2}\xi ^{2}\pm \sqrt{\left[ \left( c-\delta \right) ^{2}-D^{2}\xi ^{2}\right] ^{2}+16D^{2}K^{2}}\right\} ,
\end{equation}
 where \begin{equation}
\label{qchi}
q\equiv k\mp \delta \cdot \omega ,\, \, \, \, \xi \equiv \omega \pm \left( c-\delta \right) /D,
\end{equation}
 In the definitions (\ref{qchi}) of the shifted propagation constant and frequency,
the upper and lower signs pertain, respectively, to the (\( u,\phi  \)) and
(\( v,\psi  \)) subsystems, while in the dispersion relation (\ref{q}) the
two different signs yield two different branches of the dispersion curve, see
Fig. 1.

It follows from Eq. (\ref{q}) that the spectrum shown in Fig. 1 always has
a \emph{gap}, \begin{equation}
\label{gap}
0<-\, 4Dq<\sqrt{\left( c-\delta \right) ^{4}+16D^{2}K^{2}}-\left( c-\delta \right) ^{2},
\end{equation}
 inside which, following the general principles \cite{Martijn}, one may expect
the existence of a family of \textit{gap solitons} (note that the full dissipationless
version of the present model, including the nonlinear terms in Eqs. (\ref{eq1})
and (\ref{eq2}), is definitely nonintegrable, therefore {}``solitons{}''
are meant here simply as solitary waves). However, the objective of this work
is not to study that possible family, but to focus on the search for stable
pulses in the full model, including the gain, filtering and loss, which is much
more relevant to applications.

\subsection{Solitary-pulse solutions}

If the field is launched into one channel only, the system (\ref{eq1})-(\ref{eq4})
reduces to a subsystem of Eqs. (\ref{eq1}) and (\ref{eq3}), or (\ref{eq2})
and (\ref{eq4}), each including a CGL equation linearly coupled to its linear
dissipative counterpart. If, additionally, there is no difference in the group-velocity
mismatch in the active and passive core (i.e., \( c=\delta  \)), then, in the
reference frame moving with the common group velocity, each subsystem (\ref{eq1}),
(\ref{eq3}) or (\ref{eq2}), (\ref{eq4}), decoupled from the other one, possesses
its own pair of exact analytical solutions for chirped SPs, which were actually
found in Ref. \cite{pla1}. In the reference frame moving with the common inverse
velocity \( c=\delta  \), the exact solution takes the form

\begin{eqnarray}
u & = & A\exp (ikz)\left[ \mathrm{sech}(\eta t)\right] ^{1+i\mu },\label{eq5} \\
\phi  & = & AK(i\Gamma -k)^{-1}\exp (ikz)\left[ \mathrm{sech}(\eta t)\right] ^{1+i\mu },\label{eq6} 
\end{eqnarray}
 where the definition of the retarded time \( t \) is adjusted to the above-mentioned
moving reference frame, and the chirp coefficient is

\begin{equation}
\label{eq7}
\mu =-\frac{3}{4}D+\frac{1}{4}\sqrt{32+9D^{2}}.
\end{equation}
 The SP's inverse width \( \eta  \) and squared amplitude \( A^{2} \) (which
is the peak power, in the application to optics) are expressed in terms of the
wavenumber \( k \),

\begin{eqnarray}
\eta ^{2} & = & \frac{2k\left( \Gamma -1\right) }{k\left( D\mu -2\right) +\Gamma \left[ \left( 1-\mu ^{2}\right) D+4\mu \right] },\nonumber \\
 &  & \nonumber \\
A^{2} & = & \left( 3+\frac{1}{4}D^{2}\right) \mu \eta ^{2},\label{eq9} 
\end{eqnarray}
 and, finally, the wavenumber itself is determined by a cubic equation,

\begin{equation}
\label{eq10}
\left( \mu D-2\right) (k^{2}-1)k+\left( \left( 1-\mu ^{2}\right) D+4\mu \right) \left( k^{2}-\Gamma \left( K^{2}-\Gamma ^{2}\right) \right) =0.
\end{equation}

Clearly, physical solutions of Eq. (\ref{eq10}) are those which yield a real
wavenumber \( k \) and \( \eta ^{2}>0 \). Physical solutions usually exist
in pairs, and only the one with a larger value of the peak power may be stable.
As it was demonstrated in Refs. \cite{ref10,pla1,ref11}, the SP solution with
the larger amplitude is indeed stable in a fairly vast region in the parameter
space \( \left( \Gamma ,K,D\right)  \). These pulses definitely remain stable
in the framework of the full system (\ref{eq1}) - (\ref{eq4}). Indeed, because
the two above-mentioned subsystems (\ref{eq1}), (\ref{eq3}) and (\ref{eq2}),
(\ref{eq4}) are coupled solely by the nonlinear XPM terms, the only additional
stability condition for a pulse belonging to either subsystem is the linear
stability of the zero solution in the mate subsystem, which is always the first
condition imposed on parameters of the eligible model.

Note that the exact SP solution displayed above can be extended to the full
system of the four equations (\ref{eq1})-(\ref{eq4}) if the group-velocity
differences vanish, i.e., \( c=\delta =0 \) : the expressions (\ref{eq5})-(\ref{eq10})
then yield a solution to the system of the four equations after the transformation
\( u,v\rightarrow (u,v)/\sqrt{3} \), \( \phi ,\psi \rightarrow (\phi ,\psi )/\sqrt{3} \),
and setting \( u=v \) and \( \phi =\psi  \).

In the most general case, \( c\neq \delta  \), no exact solution for SPs is
available, but pulses can be found numerically, see below. In any case, pulses
generated by the decoupled subsystems (\ref{eq1}), (\ref{eq3}) and (\ref{eq2}),
(\ref{eq4}) move at different velocities, hence they may collide. The strong
XPM-induced nonlinear coupling between the channels, together with the dissipative
character of the system (\ref{eq1})-(\ref{eq4}), may give rise to complex
dynamical behavior as a result of the collisions.

Our objective in this work is to study in detail collisions between SPs in the
system (\ref{eq1})-(\ref{eq4}) and their stability. Note, in particular, that
in the case when the inverse-group-velocity differences in the active and passive
cores are large and nearly equal, i.e., \( c\approx \delta \gg 1 \), the subsystems
(\ref{eq1}), (\ref{eq3}) and (\ref{eq2}), (\ref{eq4}) nearly decouple, therefore
quasi-elastic collisions are expected in this case, while at smaller values
of the group-velocity mismatch collisions may be strongly inelastic. These expectations
are corroborated by numerical simulations which are displayed below.

\section{An analytical approach to collisions between solitary pulses}

\subsection{The perturbation theory}

In the cases of practical interest to fiber-optic telecommunications, the model
is far from any exactly integrable limit, therefore only direct numerical simulations
of collisions between pulses (and of their stability), results of which will
be summarized in the next section, are really relevant. Nevertheless, some qualitative
insight into the collision problem can be gained from an analytical approach,
assuming that pulses may be approximated as quasi-solitons. Within the framework
of such an approach in its most general possible form, each pulse, in its own
reference frame (in which an exact solution is given by Eqs. (\ref{eq5}) -
(\ref{eq10}), assuming that \( \delta =c \)), is taken as \begin{eqnarray}
u & = & A\, f(\eta (t-T))\exp (ikz-i\omega t)\, ,\label{upert} \\
\phi  & = & \Phi \, g(\eta (t-T))\exp (ikz-i\omega t)\, ,\label{phipert} \\
v & = & \, A\, f(\eta (t+T))\exp (ikz+i\omega t),\label{vpert} \\
\psi  & = & \Phi \, g(\eta (t+T))\exp (ikz+i\omega t).\label{psipert} 
\end{eqnarray}
 Here \( f(\eta t) \) and \( g(\eta T) \) are (generally speaking, complex)
functions accounting for a particular shape of the unperturbed pulses, \( \eta  \)
being their inverse temporal width, \( A \) and \( \Phi  \) are amplitudes
of their two components, and \( \pm \omega  \) and \( \pm T \) are shifts
of the pulses' central frequencies and temporal positions due to the interaction
between them (we consider the interaction between identical pulses, hence the
symmetry between the expressions (\ref{upert}), (\ref{phipert}) and (\ref{vpert}),
(\ref{psipert})). Each component of the pulse has its own effective mass, for
instance, \begin{equation}
\label{M}
M_{u}=A^{2}\eta ^{-1}\int _{-\infty }^{+\infty }\left| f(x)\right| ^{2}dx,\, \, M_{\phi }=\Phi ^{2}\eta ^{-1}\int _{-\infty }^{+\infty }\left| g(x)\right| ^{2}dx
\end{equation}
 (note that, in the absence of losses and gain, \( M_{u}+M_{\phi } \), as well
as \( M_{v}+M_{\psi } \), are the conserved optical energies in each subsystem).
For the pulse given by the solution (\ref{eq5}) - (\ref{eq10}), one finds
\begin{equation}
\label{Mparticular}
M_{u}=2\eta ^{-1}A^{2},\, \, M_{\phi }=M_{u}K^{2}/\left( k^{2}+\Gamma ^{2}\right) .
\end{equation}

The XPM-induced coupling between the two subsystems gives rise to a potential
force of attraction between the pulses, which can be calculated by means of
well-known methods (see, e.g., Refs. \cite{ref13,ref14}), provided that XPM
may be treated as a small perturbation (conditions for applicability of this
assumption will be considered below). It is also known that the filtering term
in Eqs. (\ref{eq1}) and (\ref{eq2}) gives rise to an effective friction force,
which, in the most general case, can be evaluated and combined with the potential
force by means of the balance equation for the pulse's momentum (as it was done,
for instance, in Refs. \cite{me}). As a result, one arrives at evolution equations
for the soliton's position and frequency shifts in the following general form,
\begin{eqnarray}
\frac{dT}{dz} & = & -D\omega ,\label{T} \\
 &  & \nonumber \\
\frac{d\omega }{dz} & = & -\kappa \eta ^{2}\frac{M_{u}}{M_{u}+M_{\phi }}\omega -\frac{A^{4}}{\left( M_{u}+M_{\phi }\right) }U\, ^{\prime }(\eta \left( T+cz\right) )\, ,\label{omega} 
\end{eqnarray}
 where the prime stands for the derivative, and the friction coefficient and
interaction potential are \begin{eqnarray}
\kappa  & = & \frac{\int _{-\infty }^{+\infty }\left| df(x)/dx\right| ^{2}dx}{\int _{-\infty }^{+\infty }\left| f(x)\right| ^{2}dx}\, ,\label{kappa} \\
 &  & \nonumber \\
U(y) & = & \int _{-\infty }^{+\infty }\left| f(x-y)\right| ^{2}\left| f(x+y)\right| ^{2}dx\, \label{U} 
\end{eqnarray}
 (the XPM coefficient \( \sigma  \) was set equal to its physical value \( 2 \)).
The additional term \( cz \) in the argument of the potential in Eq. (\ref{omega})
is generated by the group-velocity difference between the two channels, and
the ratio of the masses in the friction term on the right-hand side of Eq. (\ref{omega})
appears since the friction force acts only on the \( u \)-component of the
pulse, but not on its \( \phi  \)-component. Note that these general equations
are also valid in the case of dispersion management, when \( D \) is not a
constant, but a function of \( z \) \cite{me}.

For the pulses with the shape given by Eqs. (\ref{eq5}) - (\ref{eq10}), one
can find, from the expressions (\ref{kappa}) and (\ref{U}), that \begin{eqnarray}
\kappa  & = & \left( 4/3\right) \left( 1+\mu ^{2}\right) ,\, \, \frac{M_{u}}{M_{u}+M_{\phi }}=\frac{1}{1+\displaystyle \frac{K^{2}}{k^{2}+\Gamma ^{2}}},\label{kappaparticular} \\
 &  & \nonumber \\
U(y) & = & 4\frac{2y\cosh (2y)-\sinh (2y)}{\sinh ^{3}(2y)}\, \, .\label{Uparticular} 
\end{eqnarray}
 It is worthy to mention that the expression (\ref{Uparticular}) contains no
singularity at \( y\rightarrow 0 \).

To predict results of the collisions in the general case, the nonlinear non-autonomous
(\( z \)-dependent) ODEs (ordinary differential equations) (\ref{T}) and (\ref{omega})
with the effective potential (\ref{Uparticular}) must be solved numerically.
In view of the complexity of this ODE system and its approximate character,
it makes sense to focus, instead, on direct simulations of the underlying PDEs
(partial differential equations) (\ref{eq1}) - (\ref{eq4}, which will be done
below. Nevertheless, some results can be obtained directly from ODEs (\ref{T})
and (\ref{omega}). In particular, the most essential effect observed in direct
simulations of the underlying PDEs is an inelastic outcome of the collision
(merger or complete decay of the pulses), provided that the group-velocity difference
\( c \) is below a certain critical (threshold) value \( c_{\mathrm{cr}} \).
This value may be, very roughly, estimated as that at which the friction and
potential forces in Eq. (\ref{omega}) are comparable, that yields \begin{equation}
\label{estimate}
c_{\mathrm{cr}}\sim \frac{|D|A^{2}}{\eta \left( 1+\mu ^{2}\right) }\, .
\end{equation}
 To obtain this estimate, it was set that \( M_{u}/\left( M_{u}+M_{\phi }\right) \approx 1 \),
which is true in the cases considered below, the expression (\ref{kappaparticular})
for the friction coefficient was used, and it was naturally assumed that, for
a nontrivial collision, the maximum value of the frequency shift \( \omega  \)
is on the order of \( Dc \), see Eq. (\ref{T}). Below, it will be seen that
the crude estimate (\ref{estimate}) helps to understand the fact that \( c_{\mathrm{cr}} \)
is much smaller for the case of normal dispersion than for pulses propagating
under anomalous dispersion.

\subsection{Collision-induced position shifts of the pulses}

The ODEs (\ref{T}) and (\ref{omega}) can be used to obtain \emph{quantitative}
results in the limiting case of large \( c \), so that \begin{equation}
\label{condition}
\left| \frac{dT}{dz}\right| \ll c.
\end{equation}
 In fact, this is the case when the interaction of the pulses due to XPM may
be treated as a small perturbation, and all the above approach is strictly valid.
In this case, the term \( \eta T \) in the argument of the potential in Eq.
(\ref{omega}) may be omitted, hence the equation immediately becomes linear.
Upon Substitution of \( \omega =-D^{-1}dT/dz \) from Eq. (\ref{T}) into Eq.
(\ref{omega}) and integrating once, it reduces to the following first-order
linear equation: \begin{equation}
\label{final}
\frac{d\left( \eta T\right) }{d\left( c\eta z\right) }+\kappa \frac{\eta }{c}\left( \eta T\right) =\frac{A^{2}D}{2c^{2}}U(c\eta z)
\end{equation}
 (it is more natural to consider, as final dynamical variables, the renormalized
temporal shift \( \eta T \) and propagation distance \( c\eta z \)). To obtain
Eq. (\ref{final}), it was again assumed that \( M_{u}/\left( M_{u}+M_{\phi }\right) \approx 1 \),
which will be the confirmed below, and it was substituted \( M_{u}=2A^{2}/\eta  \),
as per Eq. (\ref{Mparticular}).

Equation (\ref{final}) can be further simplified if, in addition to the condition
(\ref{condition}), the group-velocity difference between the channels is large
enough in comparison with an effective friction force, so that \begin{equation}
\label{condition2}
\kappa \eta \ll c.
\end{equation}
 In fact, this condition turns out to be less restrictive than the one (\ref{condition}),
see below. Neglecting the friction term, Eq. (\ref{final}) takes the form \begin{equation}
\label{simplest}
\frac{d\left( \eta T\right) }{d\left( c\eta z\right) }=\frac{A^{2}D}{2c^{2}}U(c\eta z)
\end{equation}

Equations (\ref{final}) and (\ref{simplest}) show that the collision is elastic
in the present case, as the inverse-group-velocity shift \( dT/dz \) is zero
at \( z=\pm \infty  \), i.e., both before and after the collision. Nevertheless,
the result of the collision is not trivial. Indeed, Eq. (\ref{simplest}) can
be used to evaluate an important characteristic of the elastic collision, viz.,
a residual temporal (position) shift of the pulse, \begin{equation}
\label{shift}
\eta \Delta T\equiv \eta \left[ T(z=+\infty )-T(z=-\infty )\right] =\frac{A^{2}D}{2c^{2}}\int _{-\infty }^{+\infty }U(x)dx\, .
\end{equation}
 This shift is important as it gives rise to the collision-induced temporal
\emph{jitter} of the pulses, see, e.g., Refs. \cite{ref13} and \cite{ref14}.
In particular, for the potential (\ref{Uparticular}) one has \( \int _{-\infty }^{+\infty }U(x)dx=2 \),
hence \begin{equation}
\label{finalshift}
\eta \Delta T=A^{2}D/c^{2}\, .
\end{equation}
 This analytical prediction will be compared below with results of direct simulations.

To conclude the analytical consideration, we note that the general condition
(\ref{condition}) takes a simple form in terms of \( \Delta T \). Indeed,
in the present case the characteristic collision distance \( \Delta z \) is
determined by the pulse's temporal width \( 1/\eta  \), so that \( \Delta z\sim 1/\left( \eta c\right)  \),
and \( dT/dz \) may then be estimated as \( \Delta T/\Delta z\sim \eta c\Delta T \).
Inserting this into Eq. (\ref{condition}), one arrives at a simple result:
\begin{equation}
\label{smallshift}
\eta \left| \Delta T\right| \ll 1,
\end{equation}
 which means that the linear equation (\ref{final}) applies to the description
of collisions between pulses if the resulting normalized temporal shift of the
pulse is small.

\section{Numerical analysis of collisions and stability of pulses}

\subsection{The approach to the problem}

We have employed the split-step Fourier algorithm to solve Eqs. (\ref{eq1})-(\ref{eq4})
numerically, using, as initial conditions, a superposition of separated waveforms
(\ref{eq5}) and (\ref{eq6}), which yield exact SP solutions for the two decoupled
subsystems (\ref{eq1}), (\ref{eq3}) and (\ref{eq2}), (\ref{eq4}). Thus,
the following initial configurations are used:

\begin{eqnarray}
u\left( 0,t\right)  & = & A\left[ \mathrm{sech}\left( \eta \left( t-T\right) \right) \right] ^{1+i\mu },\, \, \, \, \, \, \, \, \, \, \, \, \, \, \, \, \, \, \, \, \, \, v\left( 0,t\right) =A\left[ \mathrm{sech}\left( \eta \left( t+T\right) \right) \right] ^{1+i\mu },\label{eq11} \\
 &  & \nonumber \\
\phi \left( 0,t\right)  & = & \frac{AK}{i\Gamma -k}\left[ \mathrm{sech}\left( \eta \left( t-T\right) \right) \right] ^{1+i\mu },\, \, \, \, \, \, \, \, \psi \left( 0,t\right) =\frac{AK}{i\Gamma -k}\left[ \mathrm{sech}\left( \eta \left( t+T\right) \right) \right] ^{1+i\mu },\label{eq12} 
\end{eqnarray}
 which incorporate an initial temporal separation \( 2T \) between the pulses.
We will then be able to study both \textit{incomplete} and \textit{complete}
collisions, corresponding to \( T=0 \) and \( T\neq 0 \) respectively (i.e.,
collisions between initially overlapped and separated SPs, see, e.g., Refs.
\cite{ref13} and \cite{ref14} for the discussion of relative importance of
both types of the collisions).

As for the choice of parameters, in most cases we have used the values \( \Gamma =5 \)
and \( K=4 \), which are located almost in the center of the stability domain
of the exact SP solution to the decoupled subsystems (\ref{eq1}), (\ref{eq3})
and (\ref{eq2}), (\ref{eq4}) \cite{ref11}. For the dispersion parameter \( D \)
we have chosen the values \( D=\pm 18 \), in the anomalous- and normal-dispersion
regimes, respectively. These two values of \( D \) actually correspond to the
carrier wavelength near the zero-dispersion point in a dispersion-shifted fiber
\cite{ref11} (recall that Eqs. (\ref{eq1}) and (\ref{eq2}) are normalized
so that the effective filtering coefficients in them are set equal to \( 1 \)).
As was shown in Ref. \cite{ref11}, for a typical physically relevant value
of the filtering, the corresponding values \( |D| \) are indeed close to \( 18 \).
Also, this value of \( D \) gives rise to the best stability characteristics
for SPs in the single channel model. Using these values for \( \Gamma  \),
\( K \), and \( D \), the other parameters of the exact-SP solution can be
found from Eqs. (\ref{eq5})-(\ref{eq10}): in the case of anomalous dispersion
(\( D=+18 \)),

\begin{equation}
\label{eq13}
\mu =0.074,\, \, \, \, \, \, \eta =1.57,\, \, \, \, \, \, k=23.35,\, \, \, \, \, \, A^{2}=44.56,
\end{equation}
 and for the normal dispersion (\( D=-18 \)),

\begin{equation}
\label{eq14}
\mu =27.074,\, \, \, \, \, \, \eta =0.058,\, \, \, \, \, \, k=23.35,\, \, \, \, \, \, A^{2}=22.37.
\end{equation}
 In the anomalous-dispersion regime, the pulses are much narrower, and (quite
naturally) have a much smaller chirp than their counterparts existing in the
case of the normal dispersion. Note that, for these values of the parameters,
\( M_{\phi }\approx 0.028\cdot M_{u} \) according to Eqs. (\ref{Mparticular}),
i.e., the mass of the passive-core component of the pulse is negligible in comparison
with its active-core component's mass.

Our first objective is to study in detail all possible outcomes of collisions
of stable moving pulses. In the numerical simulations we have found that, depending
on the value of the inverse group-velocity mismatch \( c \) in the active core,
three different outcomes of the collisions occur: (a) both SPs perish ({}``decay{}'');
(b) only one SP survives the collision, while the other one is destroyed (this
outcome may also be considered as a merger of two pulses into one); (c) the
pulses undergo a quasi-elastic collision, so that both reappear unscathed after
the collision.

Formation of a true stable bound state of two solitons as a result of the collision
has never been observed in the simulations. However, it will be shown below
that, in some special cases (see Fig. 5(a)), a metastable bound state is observed,
which exists over a very long propagation distance, but finally collapses into
a single pulse.

As far as the above-mentioned outcomes (a) and (b) are concerned, it is important
to mention that, in most cases (apart from the exception corresponding to the
formation of the metastable bound state, which will be specially considered
below), we have found that the corresponding collision distances are quite short,
\( z_{\mathrm{coll}}\lesssim 60 \). Thus, in these cases, after passing the
short collision distance, there remains, at most, one pulse. Obviously, the
only outcome acceptable for applications to the optical telecommunications is
a quasi-elastic collision, when both SPs restore their shapes after the interaction.

Detailed results obtained for the incomplete and complete collisions in both
normal- and anomalous-dispersion regimes, as well as results for the stability
of isolated moving pulses, are summarized below. In all the simulations, the
XPM coupling coefficient in Eqs. (\ref{eq1}) and (\ref{eq2}) was set equal
to its physical value, \( \sigma =2 \).

\subsection{Incomplete Collisions}

The results for incomplete collisions (\( T=0 \) in Eqs. (\ref{eq11}) and
(\ref{eq12})) in the case \( c=\delta  \) are summarized in Table 1, where
the three above-mentioned possible outcomes, namely {}``Decay{}'', {}``Merger{}''
and {}``Elastic{}'', are indicated. As is shown, for both the normal- and
anomalous-dispersion regimes, there exists a critical (i.e., minimum) value
\( c_{\mathrm{cr}} \) of the velocity \( c \), above which the collision is
always elastic. More importantly, the value of \( c_{\mathrm{cr}} \) in the
case of the normal dispersion is much smaller, by a factor \( \approx 5 \),
than that for the anomalous-dispersion regime. As the smaller critical velocity
difference between the channels makes it possible to have a \emph{denser} WDM
system, this result shows that the normal-dispersion regime may have an advantage
over the more traditional, from the viewpoint of the soliton transmission \cite{ref2},
anomalous-dispersion regime. On the other hand, an advantage of the latter regime
is that, inside a given channel, the pulses forming a data-carrying stream may
be packed with a higher density, as their width is much smaller according to
Eqs. (\ref{eq13}) and (\ref{eq14}). In fact, the best approach to the enhancement
of the bit-rate of the fiber-optic telecommunication link is to use the channels
in \emph{both} the anomalous- and normal-dispersion bands.

The fact that the critical value \( c_{\mathrm{cr}} \) is much smaller in the
normal-dispersion regime can be explained by the crude estimate (\ref{estimate})
obtained above on the basis of the analytical consideration. Indeed, the ratio
of the values which the expression (\ref{estimate}) takes for the parameters
(\ref{eq13}) and (\ref{eq14}) corresponding to the anomalous and normal regimes
is very small (however, the estimate is too crude for a detailed quantitative
comparison with the numerical results).

Typical examples of the three different outcomes of the incomplete collisions
are shown in Figs. 2(a)-(c) (for \( D=-18 \)) and Figs. 3(a)-2(c) (for \( D=+18 \))
in the form of contour plots. As is readily observed, SPs in the normal-dispersion
regime are indeed much broader than those in the case of the anomalous dispersion,
in accordance with what is predicted by the analytical solution, see Eqs. (\ref{eq13})-(\ref{eq14}).
In order to test the sensitivity of \( c_{\mathrm{cr}} \) to a possible mismatch
of the interchannel inverse-group-velocity differences \( \delta  \) and \( c \)
between the two cores, we have also simulated incomplete collisions in the case
\( \delta \neq c \), viz., \( \delta =1.1c \) and \( \delta =2c \). The results
are summarized in Tables 2 and 3, respectively. It is found that, in both normal-
and anomalous-dispersion regimes, a mismatch in the values of the inverse-group-velocity
differences in the active and passive cores results in an \emph{increase} of
the critical value \( c_{\mathrm{cr}} \), which was again found to be much
larger in the case of the anomalous dispersion. As was mentioned above, for
the applications it is necessary to make the critical value \( c_{\mathrm{cr}} \)
as small as possible. The results presented here clearly show that the optimum
will be attained when \( \delta =c \), i.e., when the group-velocity differences
between the channels are the same in the active and passive cores, which is
not difficult to understand in qualitative terms. Indeed, a group-velocity mismatch
between the two cores makes it necessary for the main component (in the active
core) to {}``drag{}'' its counterpart in the passive core, which inevitably
generates additional losses through the filtering term, thus enhancing inelasticity
of collisions between the pulses, cf. the perturbative treatment of the collision
in the previous section.

\subsection{Complete Collisions}

\subsubsection{Inelastic and elastic collisions}

The results of the simulations for complete collisions (i.e., for a case of
a sufficiently large initial separation between the colliding pulses) in the
case \( \delta =c \) are summarized in Table 4. In the simulations, the initial
temporal separation between the solitons is taken to be equal, approximately,
to five pulse widths, i.e., \( \eta T=2.34 \) (see Eqs. (\ref{eq5}) and (\ref{eq6})).
According to Eqs. (\ref{eq13}) and (\ref{eq14}), this choice implies \( T=40 \)
and \( T=1.5 \), in the regions of normal and anomalous dispersion, respectively.

The results shown in Table 4 suggest that, contrary to the case of incomplete
collisions, in the normal-dispersion regime the three outcomes, {}``decay{}'',
{}``merger{}'' and {}``elastic{}'', alternate with the increase of \( c \),
up to the threshold value \( c_{\mathrm{cr}}=13.92 \), past which only elastic
collisions take place. Notice, in particular, a small interval, \( 7.85\leq c<8.09 \),
where the pulses undergo elastic collisions, which is found \emph{between} the
regions where at least one pulse disappears after the collisions. Typical examples
of the {}``merger{}'', {}``decay{}'' and {}``elastic{}'' outcomes for
\( D=-18 \) are shown in Figs. 4(a) through 4(c). Notice that the collision
distance in all the cases does not exceed \( z\sim 5 \).

On the other hand, in the anomalous dispersion regime, outcomes of complete
collisions resemble what was observed in the case of the incomplete collisions:
there are only three, relatively broad intervals of \( c \), where the outcomes
are {}``merger{}'', {}``decay{}'' and {}``elastic{}'', with a value of
\( c_{\mathrm{cr}} \) slightly smaller than that in the normal-dispersion regime.
However, in the case of \( D=+18 \) it is important to note that, in the {}``merger{}''
interval (\( 0\leq c<3.33 \)), the pulses demonstrate a behavior which is remarkably
different from that observed in the case of the {}``decay{}'' or {}``merger{}''
outcomes of incomplete or complete collisions in all the other cases. In this
interval, the pulses stick together and propagate in such a quasi-bound state
over a long distance (up to \( z\, \sim 3000 \)), after which one of the pulses
eventually decays. This behavior is demonstrated in Fig. 5(a) for \( c=2 \),
where the colliding SPs form the locked configuration immediately (see the inset
in Fig. 5(a)), and then they propagate, keeping this configuration, up to \( z\sim 3000 \),
where the merger eventually takes place through the destruction of one of the
pulses. Examples of other outcomes of the collision, namely, decay of both pulses
and their elastic collision, are shown in Figs. 5(b) and 5(c), respectively,
with the collision distance being very small, \( z\sim 0.1 \).

\subsubsection{Position shifts of the pulses in the case of elastic collisions}

A significant post-collision effect, in the case of the complete elastic collisions
between the SPs, is a temporal shift \( \Delta T \). The shift is apparent,
for instance, in Figs. 4(c) and 5(c). We have performed simulations to obtain
the normalized temporal shift \( \eta \Delta T \) as a function of the inverse-group-velocity
difference \( c=\delta  \), for different initial separations between SPs.
The results are shown in Fig. 6 (for \( D=-18 \)) and Fig. 7 (for \( D=+18 \));
note that these figures display the region of the values of \( c \) which is
relevant to the applications, according to the estimate (\ref{20}). Each curve
starts from the maximum value of the temporal shift corresponding to the critical
value \( c=c_{\mathrm{cr}} \). As is seen, the smallest value of the temporal
shift is attained in the case of the normal dispersion. Also, it is observed
that a larger initial separation \( \eta T \) leads to smaller values of the
temporal shift in the case of the normal dispersion, while the opposite holds
in the anomalous-dispersion case.

It is quite relevant to compare the numerical results displayed in Figs. 6 and
7 with the analytical prediction (\ref{finalshift}), which should be relevant
under the conditions (\ref{smallshift}) and (\ref{condition2}). First of all,
the substitution of the expression (\ref{finalshift}) into the condition (\ref{smallshift})
leads to an inequality \( c^{2}\gg A^{2}|D| \). The values of the parameters
being taken as per Eqs. (\ref{eq13}) and (\ref{eq14}), we conclude that this
inequality holds in the parts of Figs. 6 and 7 where irregular oscillations
go over into a systematic decay of \( \eta \Delta T \) with the increase of
\( c \).

Next, using the expression (\ref{kappaparticular}) for the effective friction
coefficient which appears in the second condition (\ref{condition2}), and the
same values of the parameters from Eqs. (\ref{eq13}) and (\ref{eq14}), one
sees that this condition readily holds in the case of the anomalous dispersion,
shown in Fig. 7, and in the case of normal dispersion (Fig. 6) it holds for
large values of \( c \), where the above-mentioned systematic decay of \( \eta \Delta T \)
takes place.

Thus, one may expect that the analytical result (\ref{finalshift}) may be correct
for sufficiently large values of \( c \) in both cases. The inspection of Figs.
6 and 7 corroborates this expectation: despite a considerable scatter of the
values of the normalized shift, depending on the initial value of the temporal
separation between the pulses, the analytically predicted dependence (\ref{finalshift})
not only qualitatively agrees with the numerical results for large \( c \),
but also, as one can readily check, numerical values of the shift, as predicted
analytically and found from the simulations, are fairly close.

\subsection{The full-stability region for the solitary pulses}

The results presented above were restricted to the fixed values of the loss
and coupling coefficients, \( \Gamma =5 \) and \( K=4 \). It is also important
to investigate pulse collisions in the full parameter space (\( \Gamma ,K,c \))
, and, in particular, to identify parametric domains where solely elastic complete
collisions occur. These domains actually represent regions of the \emph{full
stability} of SPs in the system of Eqs. (\ref{eq1})-(\ref{eq4}), since collisions
are natural finite perturbations in this model, against which the pulses must
be stable, as well as against infinitesimal perturbations (in this extended
definition of the stability, we take into regard only complete collisions, which
are inherent perturbations, and disregard incomplete collisions, that strongly
depend upon particular initial conditions).

The thus defined full-stability domains, found by means of systematic simulations,
are displayed in Figs. 8 and 9 for the cases of the normal and anomalous dispersion,
respectively, as gray regions in the (\( \Gamma ,K \)) parametric plane for
different values of the group-velocity parameter within \( 0\leq c\leq 10 \).
Notice that the domains have, roughly speaking, a boomerang-like shape, resembling
the corresponding domain found earlier in the single-channel stabilized CGL
model, described by the decoupled subsystems (\ref{eq1}), (\ref{eq3}) or (\ref{eq2}),
(\ref{eq4}) \cite{ref11}. More importantly, in both cases \( D=\pm 18 \),
there exists a minimum value of \( c \) necessary for the collisions to be
elastic, which is significantly lower in the case of the anomalous dispersion,
namely \( c_{\mathrm{min}}=2 \) for \( D=+18 \), and \( c_{\mathrm{min}}=6 \)
for \( D=-18 \). It should also be noted that the unshaded triangular region
shown in Fig. 9 corresponds to the case where the two pulses co-propagate undistorted
without actual interaction over a very long distance (\( z\approx 4000 \)).

\section{The three-channel model}

The model considered in this work can be extended to explicitly include a larger
number of WDM channels, i.e., a larger number of the CGL equations, each being
coupled to its linear dissipative counterpart. The simplest generalization contains
three channels, which are described by the following system of the coupled CGL
equations: \begin{eqnarray}
i\left( u_{z}+2cu_{t}\right) +\left( \frac{1}{2}D-i\right) u_{tt}-iu+\left( |u|^{2}+2|v|^{2}+2|w|^{2}\right) u & = & K\phi ,\label{eq15} \\
iw_{z}+\left( \frac{1}{2}D-i\right) w_{tt}-iw+\left( |w|^{2}+2|u|^{2}+2|v|^{2}\right) w & = & K\chi ,\label{eq16} \\
i\left( v_{z}-2cv_{t}\right) +\left( \frac{1}{2}D-i\right) v_{tt}-iv+\left( |v|^{2}+2|u|^{2}+2|w|^{2}\right) v & = & K\psi ,\label{eq17} \\
i\left( \phi _{z}+2c\phi _{t}\right) +i\Gamma \phi  & = & Ku,\label{eq18} \\
i\chi _{z}+i\Gamma \chi  & = & Kw,\label{eq19} \\
i\left( \psi _{z}-2c\psi _{t}\right) +i\Gamma \psi  & = & Kv,\label{eq20} 
\end{eqnarray}
 where we have assumed that each channel has the same group velocity in the
active and passive cores, while the relative velocity between adjacent channels
is \( 2c \), as in the above model (\ref{eq1})-(\ref{eq4}). Although detailed
study of the extended model is beyond the scope of this work, we give here an
example of the existence of \emph{fully stable} pulses in the system (\ref{eq15})-(\ref{eq20}),
which undergo elastic complete collisions with each other. As is shown in Fig.
10 (for \( \Gamma =5 \), \( K=4 \), \( c=20 \), and \( \eta T=2.34 \)),
the three pulses after traveling a distance of \( z\approx 1 \) collide and
then restore their shapes and propagate undistorted. Thus, we may conjecture
that the proposed stabilized scheme may be generalized to include a larger number
of channels in which case stable pulses may still experience elastic collisions.

\section{Conclusion}

In this paper we have studied in detail collisions of chirped solitary pulses
in nonlinearly coupled cubic Ginzburg-Landau (CGL) equations, each being linked
to a stabilizing dissipative linear equation. Primarily, the two-channel model
was considered. The model may be realized as wavelength-separated data-transmission
channels co-existing in a nonlinear dual-core optical fiber, which contains
an active core with gain, and a passive core, where the propagation is governed
by the linear dissipative equations. Each channel has its components in the
active and passive cores, with a linear coupling between them. Nonlinear interaction
between different channels is induced by the cross-phase modulation (XPM), which
acts in the active core only. Thus, the model describes a WDM multi-channel
fiber-optic transmission system, provided that the dispersion length of the
pulses is essentially larger than the amplification and filtering spacing, so
that the system may be considered in the approximation which assumes a uniformly
distributed bandwidth-limited gain in the active core (and a continuous passive
core, which, in reality, may consist of short segments periodically inserted
into the long fiber-optic link, together with amplifiers and filters).

If the CGL subsystems are decoupled, they possess stable chirped pulse solutions,
which can be found in an exact analytical form, provided that the group-velocity
parameter is identical in the cubic and linear equations. The XPM-induced nonlinear
coupling between the subsystems gives rise to interactions when the pulses belonging
to the different subsystems collide. By means of direct simulations, we have
studied incomplete and complete collisions in detail. Three different possible
outcomes of the collision have been found, in the cases when the dispersion
in the active core is anomalous or normal: destruction of both pulses, destruction
of one of them, and a quasi-elastic collision. In the latter case, both pulses
reappear unscathed after the collision (with some positional shifts), provided
that the group-velocity difference between the channels exceeds a critical value.
As a result, regions in the model's parameter space have been identified where
the pulses are stable against both small perturbations and mutual (complete)
collisions. An analytical perturbation theory was developed to predict the positional
shifts. The analytical results agree well with the numerical ones in the cases
when applicability conditions for the perturbation theory hold.

For WDM applications, it is important not only to guarantee the quasi-elastic
character of the collisions between pulses belonging to different channels,
but also to have the critical group-velocity difference between adjacent channels,
necessary for the elasticity of the collisions, as small as possible, so that
the wavelength separation between the channels may be minimized. To this end,
we have found that, in the case of incomplete collisions, the normal-dispersion
regime provides an essentially smaller critical velocity, whereas in the case
of complete collisions, the critical velocities are almost equal for both signs
of the dispersion. However, the region in the parameter space where complete
collisions are always elastic is essentially larger in the case of anomalous
dispersion, and another advantage of the latter case is that the temporal width
of the pulses is much smaller. On the other hand, the residual effect of elastic
collisions, viz., the temporal shift of the pulses, which contributes to the
soliton jitter in optical communications, is weakest in the case of normal dispersion.
Actually, the best solution may be to use the channels in \emph{both} normal-dispersion
and anomalous-dispersion bands in the fiber. Finally, we have shown that the
model may be extended to include more than two WDM channels, giving rise, in
an appropriate region of the corresponding parameter space, to completely stable,
three-pulse collisions being quasi-elastic.

\newpage

{\par\centering \textbf{Figure Captions} \par}

\vspace{1cm}

Figure 1. A typical form of the dispersion curves \( q(\chi ) \) (the shifted
propagation constant and frequency are defined in Eqs. (\ref{qchi})) for the
linearized system in the dissipationless approximation. The curves are shown
for \( D=+18 \) (the case of the anomalous dispersion) and \( c=\delta  \).
For normal dispersion (\( D=-18 \)), the dispersion curves are mirror images
of those shown in this figure.\\

Figure 2. Incomplete collisions of two pulses in the normal-dispersion regime
(\( D=-18 \)). (a) Decay of both pulses, with \( c=\delta =6 \). (b) Merger
of the pulses, with \( c=\delta =9 \). (c) An elastic collision, with \( c=\delta =14 \).
\\

Figure 3. Incomplete collisions of two pulses in the anomalous-dispersion regime
(\( D=+18 \)). (a) Decay of both pulses, with \( c=\delta =40 \). (b) Merger
of the pulses, with \( c=\delta =2 \). (c) An elastic collision , with \( c=\delta =48 \).\\

Figure 4. Complete collisions of two pulses in the normal-dispersion regime
(\( D=-18 \)). (a) Merger of the pulses, with \( c=\delta =10 \). (b) Decay
of both pulses, with \( c=\delta =13 \). (c) An elastic collision, with \( c=\delta =14 \).\\

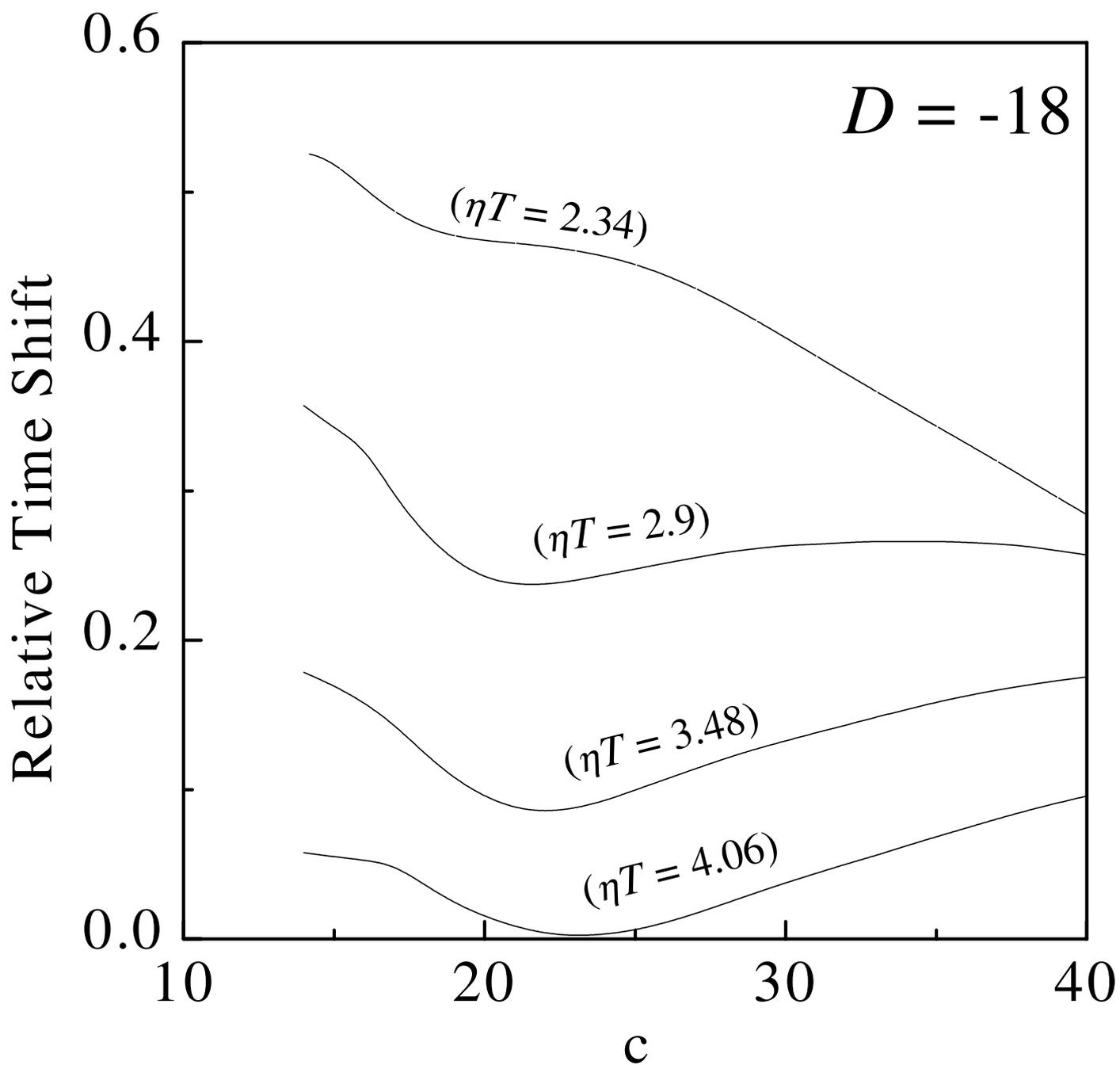
Figure 5. Complete collisions of two pulses in the anomalous-dispersion regime
(\( D=+18 \)). (a) The merger, with \( c=\delta =2 \). The two pulses get
stuck almost immediately (see the inset showing the initial stage of the collision
in detail), and then they propagate, keeping this shape up to \( z\approx 3000 \),
where the merger (in fact, destruction of one of the pulses) eventually takes
place. (b) Decay of both pulses, with \( c=\delta =5 \). (c) An elastic collision,
with \( c=\delta =48 \).\\

Figure 6. Relative temporal shift vs. the inverse-group-velocity mismatch \( c=\delta  \)
in the case of normal dispersion (\( D=-18 \)).\\

Figure 7. Relative temporal shift vs. the inverse-group-velocity mismatch \( c=\delta  \)
in the case of anomalous dispersion (\( D=+18 \)).\\

Figure 8. The full-stability region (implying the stability of the solitary
pulses against both arbitrary infinitesimal perturbations, and against collisions
with a pulse moving in the other channel) in the \( \left( K,\Gamma \right)  \)
parametric plane, at different fixed values of the inverse-group-velocity difference
between the channels, in the case of normal dispersion (\( D=-18 \)).\\

Figure 9. The same as in Fig. 8 for the case of anomalous dispersion (\( D=+18 \)).\\

Figure 10. An elastic complete collision between \emph{three} solitary pulses
in the three-channel model, in the case of normal dispersion (\( D=-18 \)),
with \( c=\delta =20 \) and \( \eta T=2.34 \). 

\newpage

{\par\centering \textbf{Table Captions} \par}

\vspace{1cm}

Table 1. Outcomes of incomplete collisions for \( \delta =c \).\\

Table 2. Outcomes of incomplete collisions for \( \delta =1.1c \).\\

Table 3. Outcomes of incomplete collisions for \( \delta =2c \).\\

Table 4. Outcomes of complete collisions for \( \delta =c \) and \( \eta T=4.7 \).

\newpage

{\centering \begin{tabular}{|c|c|c|c|}
\hline 
\multicolumn{4}{|c|}{\textbf{Incomplete Collisions (\( {\delta =c} \))}}\\
\hline 
\multicolumn{2}{|c||}{\textbf{Normal Dispersion (\( {D=-18} \))}}&
\multicolumn{2}{|c|}{\textbf{Anomalous Dispersion (\( {D=+18} \))}}\\
\hline 
\textbf{Velocity}&
 \textbf{Outcome}&
 \textbf{Velocity}&
 \textbf{Outcome}\\
\hline 
\( c<8.35 \)&
 Decay &
 \( c<3.52 \)&
 Merger \\
\hline 
\( 8.35\leq c<9.53 \)&
 Merger &
 \( 3.52\leq c<46.98 \)&
 Decay \\
\hline 
\( c>9.53 \)&
 Elastic &
 \( c\geq 46.98 \)&
 Elastic  \\
\hline 
\end{tabular}\par}

\vspace{1cm}

{\par\centering Table 1. Nistazakis, Physical Review E\par}

\newpage

{\centering \begin{tabular}{|c|c|c|c|}
\hline 
\multicolumn{4}{|c|}{\textbf{Incomplete Collisions (\( {\delta =1.1c} \))}}\\
\hline 
\multicolumn{2}{|c||}{\textbf{Normal Dispersion (\( {D=-18} \))}}&
\multicolumn{2}{|c|}{\textbf{Anomalous Dispersion (\( {D=+18} \))}}\\
\hline 
\textbf{Velocity}&
 \textbf{Outcome}&
 \textbf{Velocity}&
 \textbf{Outcome}\\
\hline 
\( c<9.88 \)&
 Decay &
 \( c<2.42 \)&
 Merger \\
\hline 
\( 9.88\leq c<10.87 \)&
 Merger &
 \( 2.42\leq c<59.29 \)&
 Decay \\
\hline 
\( 10.87\leq c<14.17 \)&
 Decay &
 \( c\geq 59.29 \)&
 Elastic \\
\hline 
\( c\geq 14.17 \)&
 Elastic &
\multicolumn{2}{|c|}{ }\\
\hline 
\end{tabular}\par}

\vspace{1cm}

{\par\centering Table 2. Nistazakis, Physical Review E\par}

\newpage

{\centering \begin{tabular}{|c|c|c|c|}
\hline 
\multicolumn{4}{|c|}{\textbf{Incomplete Collisions (\( {\delta =2c} \))}}\\
\hline 
\multicolumn{2}{|c||}{\textbf{Normal Dispersion (\( {D=-18} \))}}&
\multicolumn{2}{|c|}{\textbf{Anomalous Dispersion (\( {D=+18} \))}}\\
\hline 
\textbf{Velocity}&
 \textbf{Outcome}&
 \textbf{Velocity}&
 \textbf{Outcome}\\
\hline 
\( c<14.26 \)&
 Decay &
 \( c<1.73 \)&
 Merger \\
\hline 
\( c\geq 14.26 \)&
 Elastic &
 \( 1.73\leq c<122.17 \)&
 Decay \\
\hline 
\multicolumn{2}{|c||}{}&
 \( c\geq 122.17 \)&
 Elastic  \\
\hline 
\end{tabular}\par}

\vspace{1cm}

{\par\centering Table 3. Nistazakis, Physical Review E\par}

\newpage

\vspace{0.3cm} 

{\centering \begin{tabular}{|c|c|c|c|}
\hline 
\multicolumn{4}{|c|}{\textbf{Complete Collisions (\( {\delta =c\, \, \, and\, \, \, \eta T=4.7} \))}}\\
\hline 
\multicolumn{2}{|c||}{\textbf{Normal Dispersion (\( {D=-18} \))}}&
\multicolumn{2}{|c|}{\textbf{Anomalous Dispersion (\( {D=+18} \))}}\\
\hline 
\textbf{Velocity}&
 \textbf{Outcome}&
 \textbf{Velocity}&
 \textbf{Outcome}\\
\hline 
\( c<1.88 \)&
 Decay &
 \( c<3.33 \)&
 Merger \\
\hline 
\( 1.88\leq c<2.56 \)&
 Merger &
 \( 3.33\leq c<12.82 \)&
 Decay \\
\hline 
\( 2.56\leq c<7.85 \)&
 Decay &
 \( c\geq 12.82 \)&
 Elastic \\
\hline 
\( 7.85\leq c<8.09 \)&
 Elastic &
\multicolumn{2}{|c|}{}\\
\hline 
\( 8.09\leq c<11.71 \)&
 Merger &
\multicolumn{2}{|c|}{}\\
\hline 
\( 11.71\leq c<13.92 \)&
 Decay &
\multicolumn{2}{|c|}{}\\
\hline 
\( c\geq 13.92 \)&
 Elastic &
\multicolumn{2}{|c|}{ }\\
\hline 
\end{tabular}\par}

\vspace{0.3cm}

\vspace{1cm}

{\par\centering Table 4. Nistazakis, Physical Review E\par}
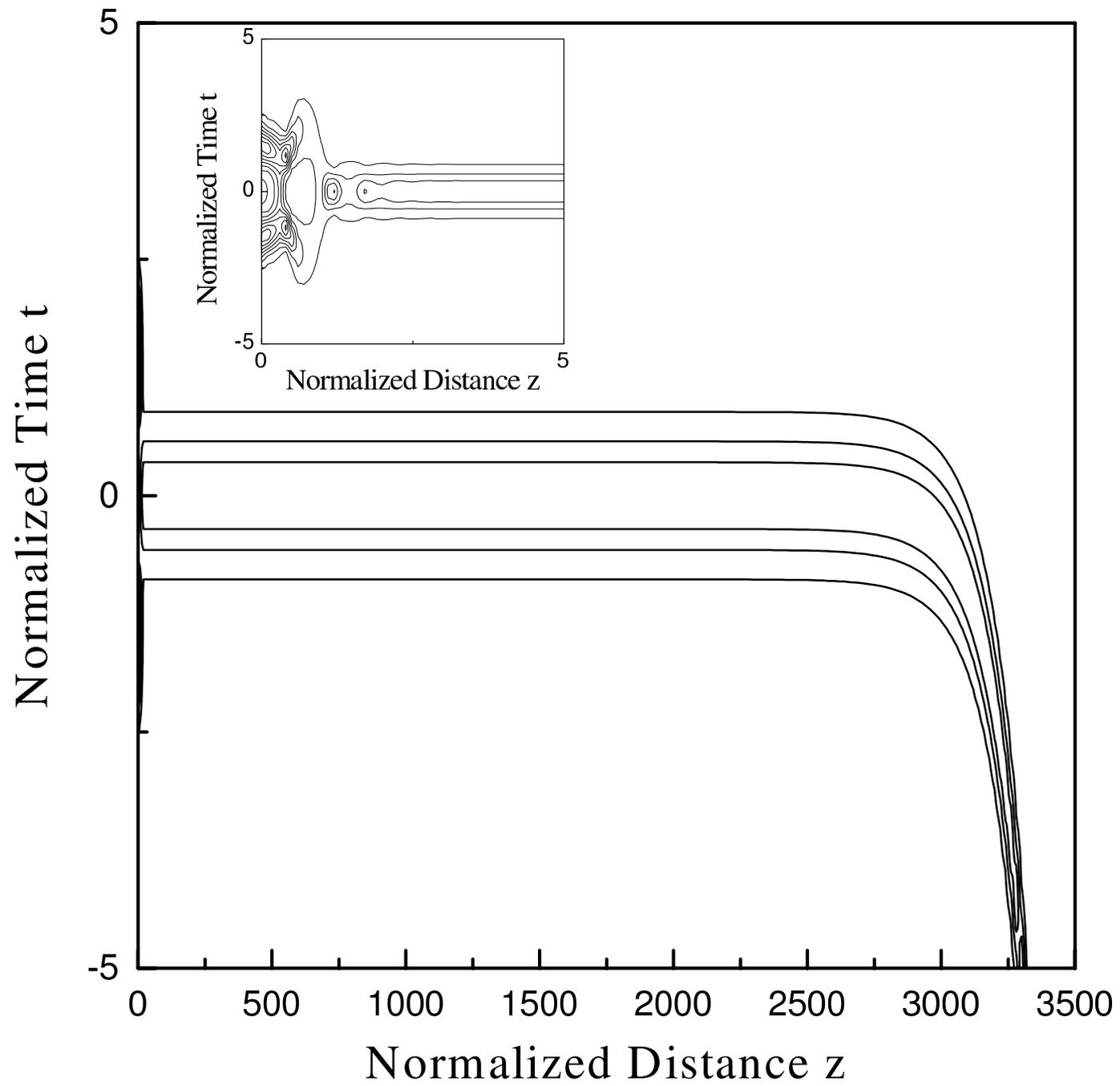
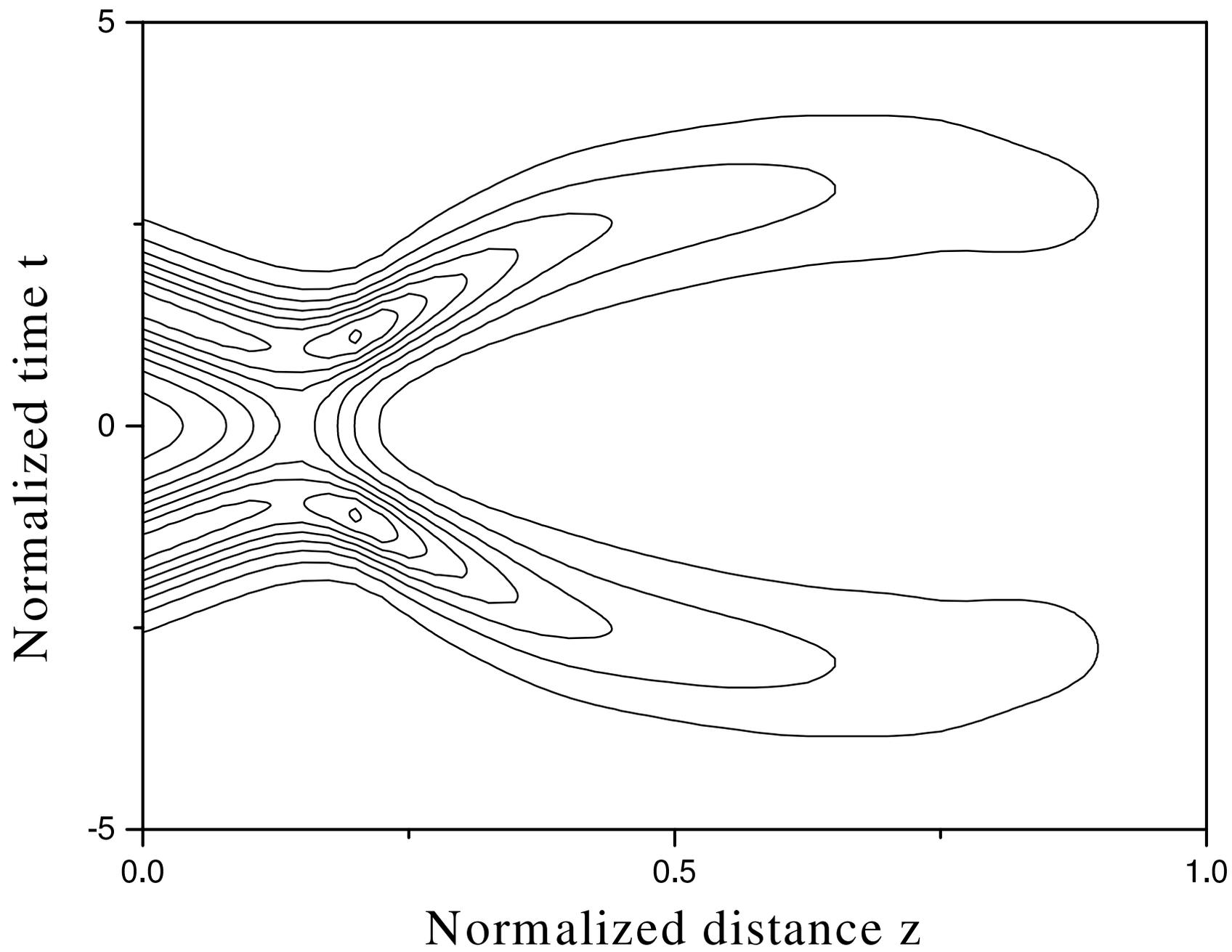
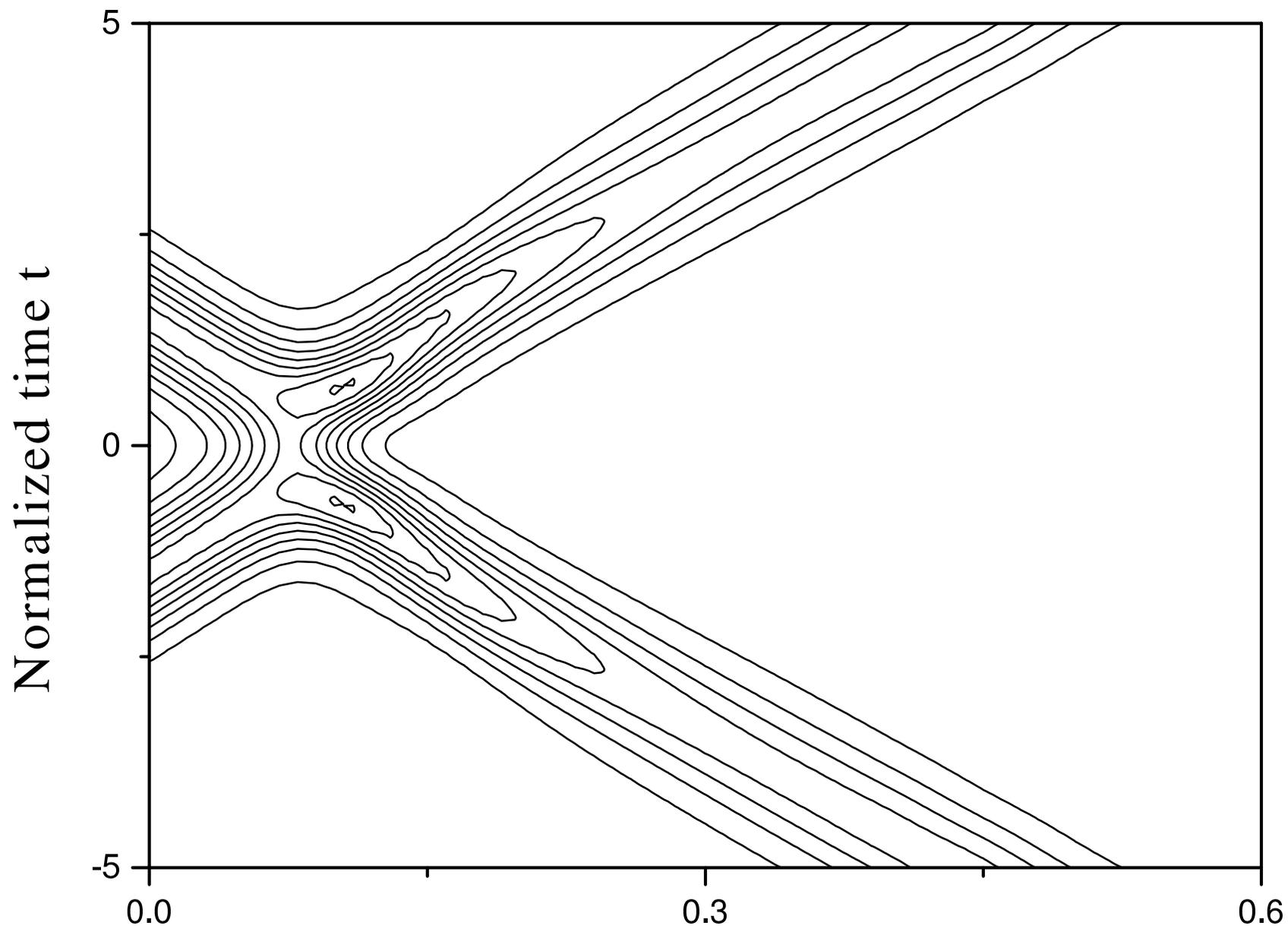
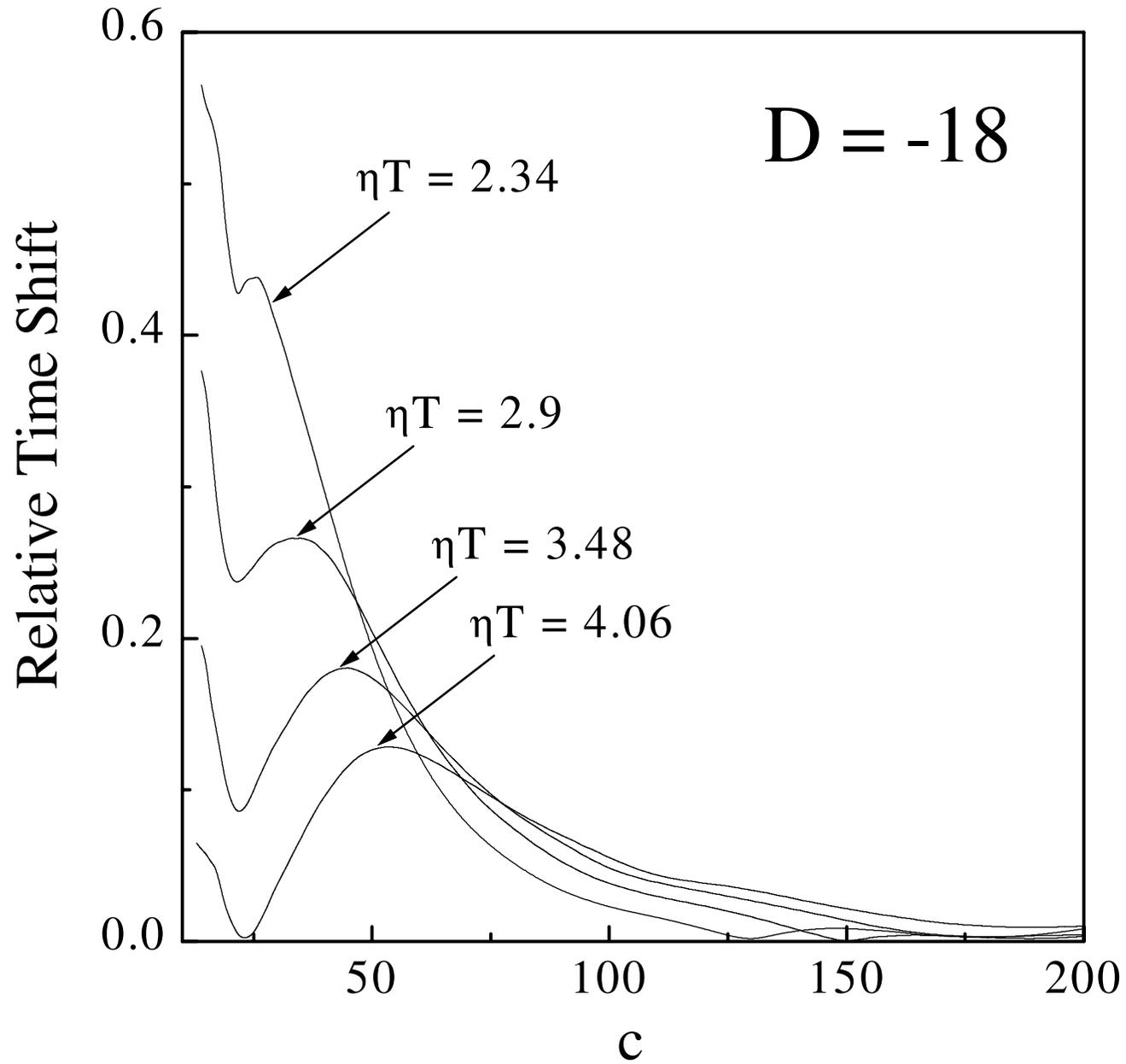
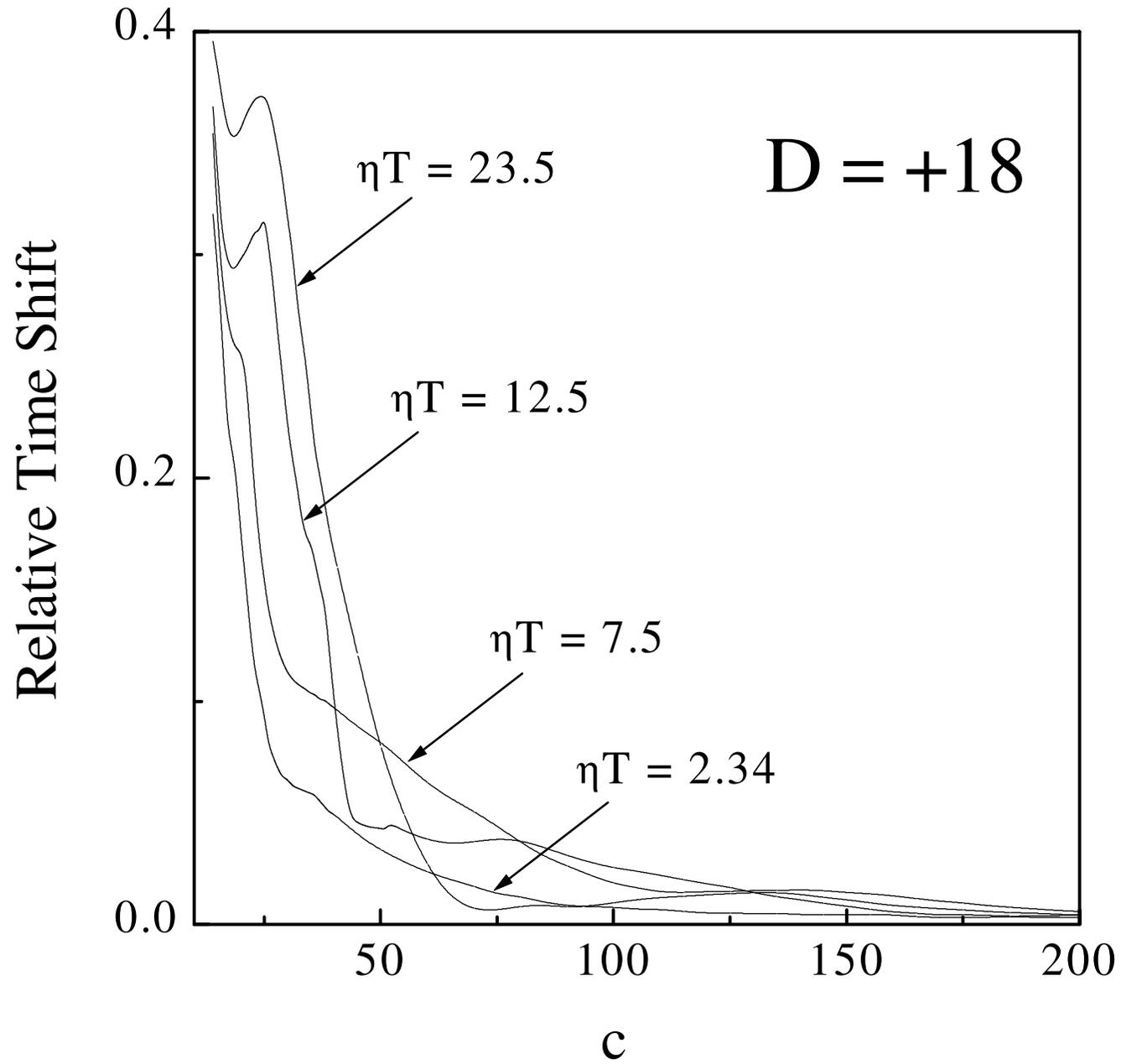
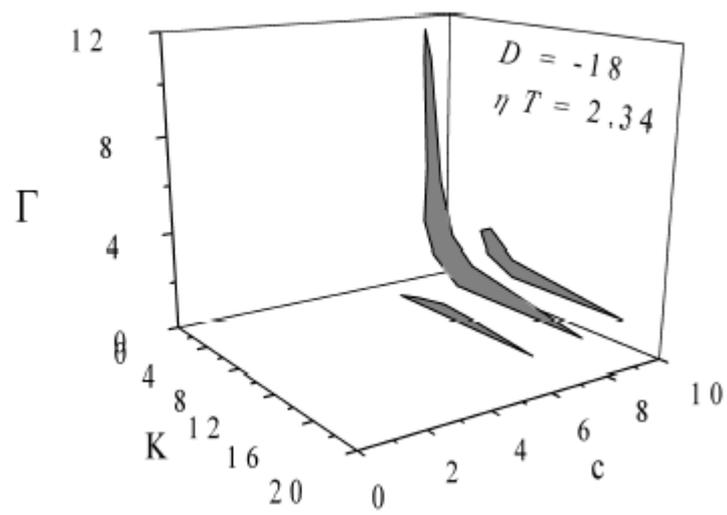
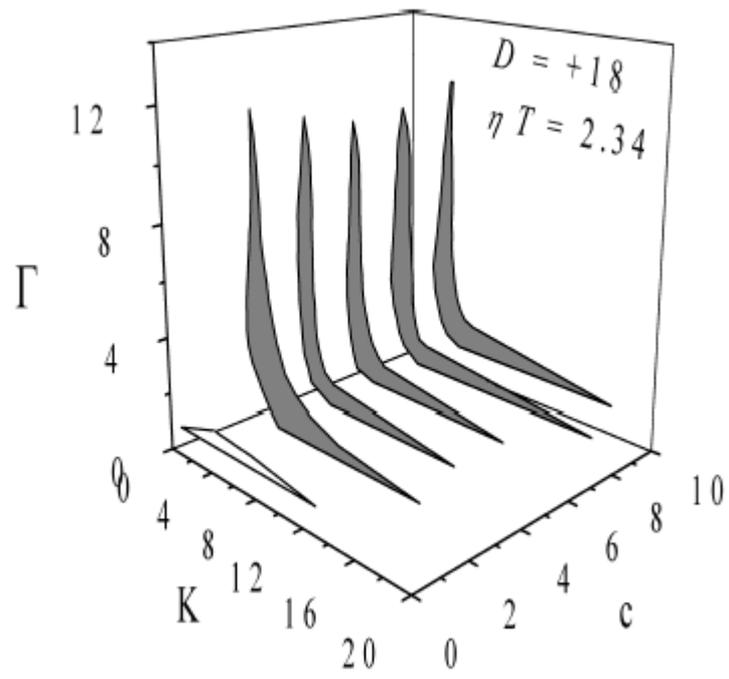
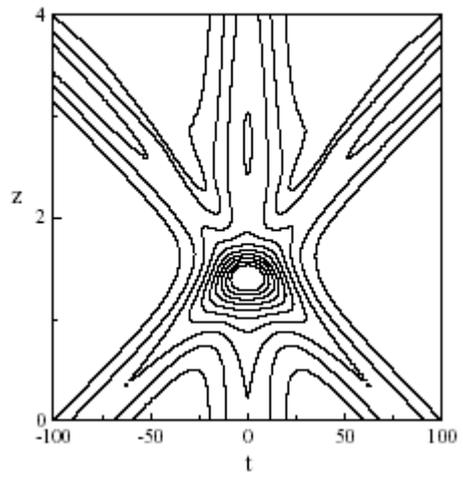
\end{document}